\documentclass[pmlr]{jmlr}

\usepackage{longtable}

\usepackage{booktabs}
\usepackage[load-configurations=version-1]{siunitx} 

\theorembodyfont{\upshape}
\theoremheaderfont{\scshape}
\theorempostheader{:}
\theoremsep{\newline}

\jmlrvolume{}
\firstpageno{1}
\editors{Sophia Sanborn, Christian Shewmake, Simone Azeglio, Nina Miolane}

\jmlrworkshop{Symmetry and Geometry in Neural Representations}

\title[Optimal packing of attractor states]{Optimal packing of attractor states in neural representations}

 \author{\Name{John J. Vastola} \Email{john\_vastola@hms.harvard.edu}\\
 \addr Department of Neurobiology, Harvard Medical School, Boston, MA, USA
 }

\begin{document}

\maketitle

\begin{abstract}
Animals' internal states reflect variables like their position in space, orientation, decisions, and motor actions---but how should these internal states be arranged? Internal states which frequently transition between one another should be close enough that transitions can happen quickly, but not so close that neural noise significantly impacts the stability of those states, and how reliably they can be encoded and decoded. In this paper, we study the problem of striking a balance between these two concerns, which we call an `optimal packing' problem since it resembles mathematical problems like sphere packing. While this problem is generally extremely difficult, we show that symmetries in environmental transition statistics imply certain symmetries of the optimal neural representations, which allows us in some cases to exactly solve for the optimal state arrangement. We focus on two toy cases: uniform transition statistics, and cyclic transition statistics. Code is available at \url{https://github.com/john-vastola/optimal-packing-neurreps23}.
\end{abstract}

\pagenumbering{arabic}  

\section{Introduction}
\label{sec:intro}

Animals' internal states appear to reflect environmental variables, like their position in space and orientation relative to some reference \citep{Barry2014,Hulse2020}, as well as their interactions with the environment, like decisions \citep{Gold2007} and motor actions \citep{Cisek2005}. As an animal acts in its environment, it must constantly update these internal states to reflect environmental changes and the results of internal computations; however, these updates cannot be \textit{instantaneous}, since biophysical limitations force internal quantities to change in a somewhat continuous fashion. They are also not \textit{error-free} due to noise in encoding, decoding, and neural dynamics \citep{Faisal2008,Haim1996}. 

How should internal states be arranged? On the one hand, an animal can act more quickly if the next relevant internal state is `near' the current one, since it can be reached more quickly. This suggests that the structure of neural representations should reflect the structure of environmental transitions; this is consistent with what is known about circuits like the head direction system, whose latent geometry mirrors the circular nature of the variable it tracks \citep{Ajabi2023}, and theoretical ideas about smoothness as a constraint on neural codes \citep{Stringer2019}. On the other hand, the closer all internal states are to one another, the easier it is for neural noise to cause problems, either via noise-induced transitions \citep{Burak2012} or by increasing the likelihood of encoding and decoding errors. In principle, a `good' arrangement of internal states strikes a balance between these two concerns: internal states must be packed closely enough that desired transitions can happen quickly, but not so closely that errors are likely. 

Given that noise sets an effective length scale for separating internal states, this issue in some ways mathematically resembles an optimal packing problem. While often quite difficult, problems like optimal sphere packing \citep{zong2008sphere} are made substantially easier to solve and understand by exploiting symmetry-related considerations. For example, Viazovska et al.'s solution of the sphere packing problem in dimensions 8 and 24 \citep{Viazovska2017,Cohn2017} crucially uses symmetry properties of the $E_8$ and Leech lattices. In this paper, we attempt to formulate a toy version of the problem of constructing an `optimal packing' of neural representations, and similarly turn to symmetry-related tools in order to say something meaningful about it. The particular symmetry-related claim we will motivate, and then use, is that an attractor-based neural representation of a Markov chain that exhibits a symmetry may also exhibit that symmetry.

\section{Mathematical formulation of optimal packing problem}
\label{sec:math}

To formalize our optimal packing problem, we need five things: a model of environment state statistics, a model of internal state transition dynamics, an encoding model, a decoding model, and a cost function. \\

\noindent \textbf{Environment dynamics.} We will model the environment as a Markov chain on $M$ states. In particular, we will assume that it can be characterized by a set of states $\mathcal{X} = \{ 1, ..., M \}$, a base probability of state occupancy $p_0(x)$ for all $x \in \mathcal{X}$, and a probability $p(y | x)$ of transitioning from any state $x \in \mathcal{X}$ to any state $y \in \mathcal{X}$ on some characteristic time scale. Since we are interested in finding representations that respect environmental transition structure \textit{when a transition occurs}, we assume without loss of generality that $p(x | x) = 0$ for all $x \in \mathcal{X}$. \\

\noindent \textbf{Internal state transition dynamics.} Let $\mathcal{Z} = \mathbb{R}^D$ (for some $D \geq M$) denote the set of all possible internal states, and assume that each environment state $x \in \mathcal{X}$ is in one-to-one correspondence with an internal attractor state $\vec{z}_x \in \mathcal{Z}$. Assume also that the positive definite matrix $\vec{\Sigma}^{-1}$ can be used to compute the distance 
\begin{equation}
D(\vec{z}_1, \vec{z}_2) := \sqrt{(\vec{z}_1 - \vec{z}_2)^T \vec{\Sigma}^{-1} (\vec{z}_1 - \vec{z}_2)}
\end{equation}
between any two internal states. The matrix $\vec{\Sigma}$ is intended to model how noisy different directions in $\mathcal{Z}$ are; different states are `closer', in the sense of being easier to reach from one another, if the line connecting them corresponds to a particularly noisy direction. 

Although it is possible to write down an extremely explicit model of internal state transition dynamics, we will consider a somewhat coarse description in order to keep our problem mathematically tractable. We will assume three things: first, that transitions are essentially between attractor basins, so that the relevant quantity is the discrete distribution $q(\vec{z}_y | \vec{z}_x)$; second, that there is a mechanism for destabilizing attractor states when a transition is desired, so that $q(\vec{z}_x | \vec{z}_x) = 0$ for all attractor states $\vec{z}_x$; and third, that transitions to good approximation only depend on the distances between states. The last assumption makes sense within a landscape picture of internal dynamics (involving $M$ attractors of similar width and depth), and can be formally justified via appealing to, e.g., Kramers' theory \citep{Kramers1940,Hanggi1990}. Explicitly, we will assume that
\begin{equation}
\begin{split}
q(\vec{z}_y | \vec{z}_x) &=  (1 - \delta_{xy}) \frac{e^{- D(\vec{z}_y, \vec{z}_x)^2/2} }{Z(\vec{z}_x)} \\
Z(\vec{z}_x) &=  \sum_{a \neq x} e^{- D(\vec{z}_a, \vec{z}_x)^2/2 } \ .
\end{split}
\end{equation}

\noindent \textbf{Encoding/decoding models.} We will assume that the encoding of an environment state $x \in \mathcal{X}$ is noisy, and that when a mistake is made, $x$ is more likely to be encoded as a state $\vec{z}_a$ \textit{near} $\vec{z}_x$. Explicitly, we will assume
\begin{equation}
p_e(\vec{z}_a | x) = \frac{\delta_{ax} e^b + (1 - \delta_{ax}) e^{- D(\vec{z}_a, \vec{z}_x)^2/2}}{e^b + Z(\vec{z}_x)}
\end{equation}
where increasing the bias $b \geq 0$ makes errors less likely. In the $b \to \infty$ limit, encoding is perfect. Assuming a uniform prior, we obtain a decoding model through Bayes' rule:
\begin{equation}
p_d(x | \vec{z}_a) = \frac{p_e(\vec{z}_a | x)}{\sum_x p_e(\vec{z}_a | x)} \ .
\end{equation}
In the $b \to \infty$ limit, since $p(\vec{z}_a | x) = \delta_{ax}$, we also have perfect decoding. \\

\noindent \textbf{Cost function.} We want to penalize different possible \textit{arrangements} of internal attractor states according to some objective function, so that optimizing that objective corresponds to identifying an optimal packing. `Optimality' here means an arrangement which, as much as possible, produces internal dynamics (i.e., movement between attractors) whose transition statistics mirror the statistics of environmental transitions (\figureref{fig:general}a). The interpretation of this is that, \textit{in the absence of any external input}, the internal state is poised to change in the same way that the environment is likely to change. 

Consider the way ring-attractor-like networks reckon with uncertainty as a concrete example of this feature: in the absence of external input, the bump representing heading direction diffuses \citep{Kutschireiter2023}, a purely internal state change that reflects the fact that moment-to-moment changes in heading direction will usually be small, and are equally likely to be clockwise or counterclockwise.

One way to formalize this desire mathematically is to ask that 
\begin{equation} \label{eq:p_int}
p_{int}(y | x) := \sum_{a, b} p_d(y | \vec{z}_b) q(\vec{z}_b | \vec{z}_a) p_e(\vec{z}_a | x) \ .
\end{equation}
on average matches $p(y | x)$, the function that determines the statistics of environmental transitions. (Equivalently: we can ask that the diagram in \figureref{fig:general}a commutes.) More precisely, we want the Kullback-Leibler divergence between $p(y | x)$ and $p_{int}(y | x)$ to be small. 

We also want to include a regularization term which enforces the fact that, all else being equal, we prefer configurations with low activity, i.e., configurations for which the norm
\begin{equation}
\Vert \vec{z}_x \Vert^2 := \vec{z}_x^T \vec{\Sigma}^{-1} \vec{z}_x
\end{equation}
is small for all attractor states $\vec{z}_x$. (This can be viewed as a kind of firing rate penalty.) Hence, we will define 
\begin{equation}
\begin{split}
J[\{ \vec{z}_x \}] :=& \ \mathbb{E}_{\vec{x}}\left\{ \ KL( p \Vert p_{int}) +  \frac{\alpha}{2} \Vert \vec{z}_x \Vert^2   \  \right\} \\
=& \ \sum_{x, y} p_0(x) p(y | x) \left\{ \ \log p(y | x) - \log p_{int}(y | x) \right\} +  \frac{\alpha}{2} \sum_x p_0(x) \Vert \vec{z}_x \Vert^2  
\end{split}
\end{equation}
as an objective over possible attractor state assignments $\{ \vec{z}_x \}$. Heuristically, we can think of the objective as representing a contest between three competing interests: low firing rate, high encoding/decoding accuracy, and internal dynamics mirroring environmental transition structure (for example, in the sense depicted in \figureref{fig:general}b). The firing rate penalty pushes all attractor states towards the origin; increasing encoding and decoding accuracy pushes all attractor states infinitely far apart; and having internal dynamics mirror environment dynamics incentivizes particular relationships between attractor states. 

As usual, we can drop terms which do not depend on the $\vec{z}_x$, so we can redefine $J$ as
\begin{equation} \label{eq:main_prob}
\begin{split}
J[\{ \vec{z}_x \}] := - \sum_{x, y} p_0(x) p(y | x) \log p_{int}(y | x) +  \frac{\alpha}{2} \sum_x p_0(x) \Vert \vec{z}_x \Vert^2   \ .
\end{split}
\end{equation}
Our central concern in the following is: \textit{under what conditions can we find optimal attractor state assignments $\{ \vec{z}_x \}$?} 

\begin{figure}[htbp]
\floatconts
  {fig:general}
  {\caption{Schematic of optimal packing problem. \textbf{a.} We want environment transition statistics $p(y | x)$ to typically match the combination of encoding, internal dynamics, and decoding, or equivalently for this diagram to commute. \textbf{b.} Intuitively, the geometric structure of the attractor landscape should match the structure of the Markov chain; for example, states with frequent transitions ought to be closer together than states between which transitions are rare.}}
  {\includegraphics[width=0.9\linewidth]{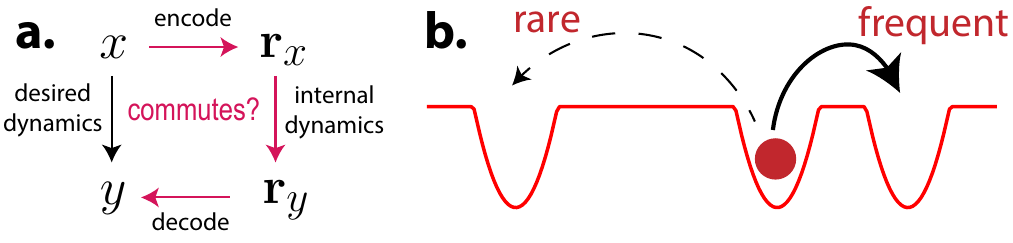}}
\end{figure}

\section{Initial observations: simplifications and symmetry}

Eq. \ref{eq:main_prob} defines a high-dimensional, nonlinear, and probably non-convex optimization problem which is in general difficult to solve. In this section, we will make several useful preliminary observations about it.\\

\noindent \textbf{Simplifying the objective.} The objective can be reparameterized in a way that makes various features of the problem clearer (see Appendix \ref{sec:app_simplify}). First, we can change variables from $\vec{z}$ to $\vec{r}$ with
\begin{equation}
D(\vec{z}_1, \vec{z}_2) = D(\vec{r}_1, \vec{r}_2) = \Vert \vec{r}_1 - \vec{r}_2 \Vert_2^2 \hspace{1in}  \vec{z}^T \vec{\Sigma}^{-1} \vec{z} = \vec{r}^T \vec{r} 
\end{equation}
by using scaled eigenvectors of $\vec{\Sigma}$ as a basis for $\mathbb{R}^D$. By doing so, we can disentangle the impact of noise anisotropy from other aspects of the problem. Let $\{ \vec{r}_x \}$ denote attractor assignments in the new coordinate system; in what follows, we will work exclusively with $\vec{r}_x$ instead of $\vec{z}_x$. Second, we can write $J$ in terms of
\begin{equation}
\langle \vec{r} \rangle := \sum_x p_0(x) \vec{r}_x \hspace{1in} d_{xy} := \Vert \vec{r}_x - \vec{r}_y \Vert_2^2 = d_{yx} \ ,
\end{equation}
i.e., the average attractor state location ($D$ scalars) and the pairwise distances between each attractor state ($M(M-1)/2$ scalars), each of which is independent of the others. There are typically less than $D M$ degrees of freedom since the problem is both rotation- and reflection-invariant, and since the objective is defined in terms of averages.\footnote{Note that $D M = (M-1) D + D \geq \frac{M(M-1)}{2} + D$. At worst, we have as many degrees of freedom as we started with, but we usually have fewer. This only works since we assumed $D \geq M$.} In terms of these variables, we have
\begin{equation} \label{eq:red_prob}
J[\{ d_{xy} \}, \langle \vec{r} \rangle] = - \sum_{x, y} p_0(x) p(y | x) \log p_{int}(y | x) + \alpha \frac{\Vert \langle \vec{r} \rangle \Vert_2^2}{2} + \alpha \sum_{a \neq b} p_0(a) p_0(b) \frac{d_{ab}^2}{4} \ .
\end{equation}
Furthermore, the optimal choice of $\langle \vec{r} \rangle$ is obvious, since it only appears in the quadratic regularization term: all optimal configurations have $\langle \vec{r} \rangle = \vec{0}$. This means that we only need to optimize the $M(M-1)/2$ pairwise distances $d_{xy}$. \\

\noindent \textbf{Symmetry.} Let $\pi: \{ 1, ..., M \} \to \{ 1, ..., M \}$ be a permutation of $\mathcal{X}$. Suppose that $\pi$ is a symmetry of the Markov chain, i.e., that
\begin{equation}
p_0(\pi(x)) = p_0(x) \hspace{1in} p(\pi(y) | \pi(x)) = p(y | x) \ .
\end{equation}
We will show that the objective function shares this symmetry. Consider the map that takes $d_{xy} \mapsto d_{\pi(x) \pi(y)}$. Because $p_{int}$ only depends on pairwise distances, we have $p_{int}(y | x) \mapsto p_{int}(\pi(y) | \pi(x))$. The relevant part of the objective becomes
\begin{equation}
\begin{split}
& \ - \sum_{x, y} p_0(x) p(y | x) \log p_{int}(\pi(y) | \pi(x)) + \alpha \sum_{a \neq b} p_0(a) p_0(b) \frac{d_{\pi(a) \pi(b)}^2}{4} \\
=& \ - \sum_{x, y} p_0(\pi(x)) p(\pi(y) | \pi(x)) \log p_{int}(\pi(y) | \pi(x)) + \alpha \sum_{a \neq b} p_0(\pi(a)) p_0(\pi(b)) \frac{d_{\pi(a) \pi(b)}^2}{4} 
 \end{split}
\end{equation}
where we have used the definition of the symmetry. But since we are summing over $x$, $y$, $a$, and $b$, it does not matter how we permute them; hence, the map we have introduced does not change the objective. 

For convex optimization problems, one can show that the unique global minimum of the objective must share its symmetries. But our problem is probably not convex, so we must settle for something weaker: in the spirit of the Purkiss principle \citep{Waterhouse1983}, we can look for solutions that share the objective's symmetries. A more rigorous analysis of Eq. \ref{eq:red_prob} may be able to show that Waterhouse's precise formulation of the Purkiss principle applies, although we do not pursue such an analysis here.

\section{Results: optimal packing for uniform and cyclic topologies}

In this section, we study the symmetric solutions of two classes of highly symmetric packing problems: the first assumes environment statistics are uniform (each state is equally likely to be next), and the second assumes statistics are cyclic (transitions occur on a ring).

\begin{figure}[htbp]
\floatconts
  {fig:uniform}
  {\caption{Solution of packing problem for a Markov chain with a uniform topology (see top right graphs). \textbf{a.} The optimal solution has the distance between all states equal, which means the geometry is that of a ($M-1$)-simplex. \textbf{b.} The objective function versus the distance $d$. Note the bifurcation as the bias decreases. \textbf{c.} The optimal distance as a function of the bias. It is zero for very small or large biases.}}
  {\includegraphics[width=\linewidth]{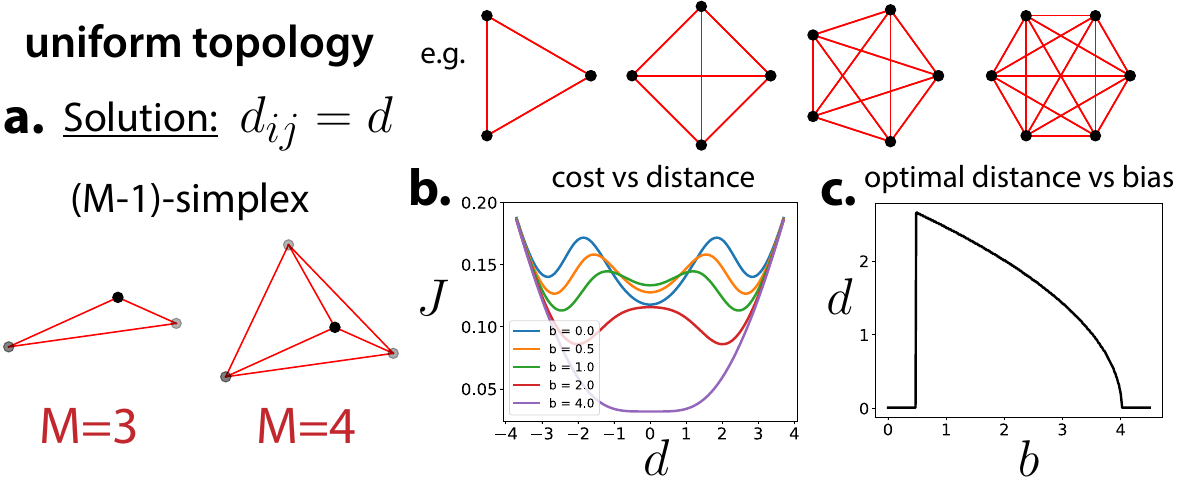}}
\end{figure} 

\subsection{Optimal packing for uniform topology}

If environmental transitions are completely uniform, then the corresponding Markov chain can be visualized as a complete graph (each vertex is connected to all others) on $M$ vertices (\figureref{fig:uniform}). Every possible permutation of $\mathcal{X}$ is a symmetry of this Markov chain, so our symmetric solution has $d_{ij} = d$ for all pairs. Hence, the geometry of this representation is completely determined: it is an ($M-1$)-simplex, an $(M-1)$-dimensional object (see \figureref{fig:uniform}a for two examples). 

After some algebra (see Appendix \ref{sec:app_uniform}), we find that the objective function can be written
\begin{equation*}
J = 2 \log\left[ e^b + (M-1) e^{-d^2/2} \right] - \log\left[ e^{2b} + 2 (M-2) e^{b - d^2/2} + (M^2 - 3 M +5) e^{-d^2}  \right] + \alpha \frac{M-1}{M} \frac{d^2}{4} 
\end{equation*}
up to unimportant additive constants. The cost $J$ as a function of the distance $d$ between attractor states is plotted in \figureref{fig:uniform}b for different values of the bias $b$. We observe fairly interesting qualitative behavior: when the bias is small, encoding and decoding are highly noisy, and the optimal solution places all states at the origin; when the bias is large, performance is good even if all states are placed arbitrarily close together; finally, when the bias is moderate, a nontrivial solution exists. In the region with a nontrivial solution, the optimal distance decreases monotonically as the bias increases (\figureref{fig:uniform}c).

\subsection{Optimally packing four attractor states}

The simplest possible cyclic Markov chain that is not uniform has $M = 4$ states. The relevant graph is a square, and the relevant symmetry group is $D_4$, the dihedral group of order $4$. Symmetry constraints (in particular, rotation symmetry) tell us that $d_{12} = d_{23} = d_{34} = d_{41} =d$, and that $d_{13} = d_{24} = L$; the precise values of $d$ and $L$ must be determined by optimizing the objective.

Naively, we might expect that the answer should be a square in neural activity space; however, this is not necessarily true, even after our correction for noise anisotropy. First off, non-square arrangements of four points exist with equal edge lengths and diagonals. For example, a square folded along one diagonal, with its angles slightly distorted, satisfies the distance constraints (\figureref{fig:square}a).

After some algebra (see Appendix \ref{sec:app_four}), we find that the objective function can be written
\begin{equation*}
\begin{split}
J &= \log Z + 2 \log(e^b + Z) + \frac{d^2}{2} - \log\left[  e^{2 b}  + 4 e^b  e^{-L^2/2} + 3 e^{  - L^2} + 4 e^{ - d^2}   \right] + \frac{\alpha}{4} \left(  \frac{d^2}{2} + \frac{L^2}{4} \right)  \ ,
\end{split}
\end{equation*}
which we plot in \figureref{fig:square}b for different values of the bias $b$. As in the uniform case, we observe phase-transition-like behavior: for small bias, the optimal solution places all states at the origin because it is too noisy; for a moderate bias, there is a nontrivial solution. Unlike before, in the case of a large bias, not all states are placed at the origin: the meaning of $d = 0$ and $L \neq 0$ is essentially that states are `glued' together so that the quadrilateral becomes a line. 

Perhaps unexpectedly, the optimal $d$ and $L$ do not produce a square representation in general, since $L/d \neq \sqrt{2}$ except at a special bias value (\figureref{fig:square}c). The actual arrangement produced (effectively three-dimensional, since four points are involved) is depicted for a few bias values in \figureref{fig:square}d. Note that if the bias is \textit{just} large enough for a nontrivial solution to exist, the arrangement is approximately square.

\begin{figure}[htbp]
\floatconts
  {fig:square}
  {\caption{Solution of packing problem for a Markov chain with a square topology. \textbf{a.} The optimal solution has two undetermined distances $d$ and $L$; depending on their ratio, non-square solutions are possible. \textbf{b.} The objective function versus $d$ and $L$ for different bias values. Note phase-transition-like behavior for small and large biases. \textbf{c.} The optimal $d$ and $L$ values generally do not have $L/d = \sqrt{2}$ when a nontrivial solution exists. \textbf{d.} The optimal arrangement is generally not square.} }
  {\includegraphics[width=\linewidth]{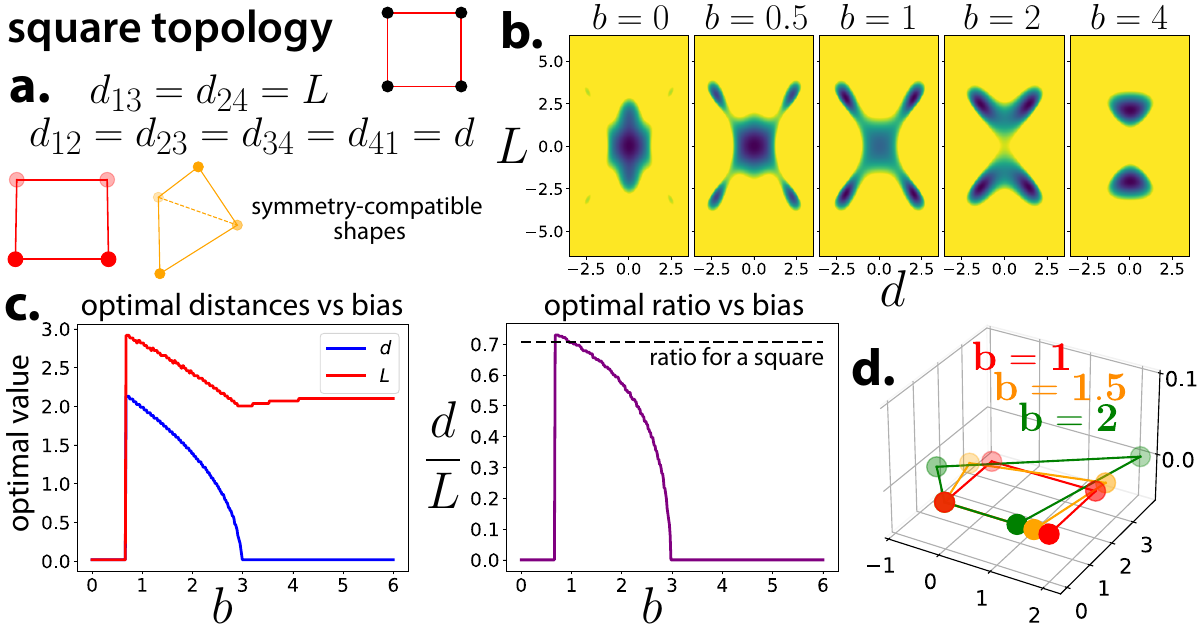}}
\end{figure}

\subsection{Optimal packing for cyclic topology}

If environmental transitions are cyclic, then the corresponding Markov chain can be visualized as a cycle graph $C_M$. By rotational symmetry, we have as many distances to determine as there are distinct vertex-vertex distances in a regular polygon (and this number differs depending on whether $M$ is even or odd). The relevant symmetry group is $D_M$, the dihedral group of order $M$. In principle, this problem could be analyzed using the same approach we used to analyze the $M = 4$ case; however, this is extremely tedious. An interesting special case that may be tractable is where the bias $b$ is large, in which case $p_{int}$ is nearly (see Appendix \ref{sec:app_cyclic})
\begin{equation*}
\begin{split}
p_{int} \approx \frac{e^{-d_1^2/2}}{Z} \left\{ 1 + 2 e^{-b} \left[ 2 \sum_{k=1}^{M/2-1} e^{-d_k^2/2 - d_{k+1}^2/2 + d_1^2/2}  -  Z \right] \right\} \hspace{0.2in} Z = e^{-d_{N}^2/2} + \sum_{k = 1}^{M/2-1} 2 e^{-d_k^2/2} \ ,
\end{split}
\end{equation*}
where $d_1$, $d_2$, and so on are the various state-state distances. Our intuition from the $M = 4$ case suggests the following: the optimal configuration should have all state-state distances (approximately) equal to those of a regular polygon near the minimal bias sufficient to support a nontrivial configuration.

\section{Discussion}

We have attempted to formulate the problem of packing attractor states in a neural representation so that internal transition statistics match environmental transition statistics as much as possible. Here, we will make comments mainly about two things: other potentially tractable Markov chain topologies, and possible generalizations of our formulation.

Other classes that may be tractable include straightforward generalizations of the cyclic topology (e.g., spherical or toroidal) and a translation-invariant lattice topology. It may be somewhat surprising that the optimization problem proved somewhat difficult \textit{even} in the case of a cyclic topology. This suggests that exact solutions may be hard to obtain, even for models with a high degree of symmetry, unless an approximation (e.g., high bias, large $M$) or special trick is used. Simulation may be a more effective route towards understanding the behavior of this problem. For example: is the global minimum generally unique? It was in the cases we examined, but this does not imply much about the general case.

At least superficially, our packing problem somewhat resembles the problem of finding Euclidean graph embeddings \citep{Cai2018}. (There are important qualitative differences in the functional form of the objective, however.) It may be possible to adapt some results from that setting to provide insight here.

A variety of generalizations are possible, both to make the problem more mathematically interesting and to make it more relevant to neuroscience. If we return to our definition of $p_{int}(y | x)$ (Eq. \ref{eq:p_int}), a few become obvious: we can use more realistic encoding/decoding models, like Poisson spiking models or probabilistic population codes \citep{Ma2006,Vastola2023}; we can use a more realistic dynamics model, like a recurrent neural network; and we can define distances or dynamics on a non-Euclidean space. It is unclear which more complex choices would still yield a somewhat tractable mathematical problem. 

An interesting phenomenon potentially related to the problem we consider here is the fact that neural representations tend to drift \citep{Driscoll2022,Masset2022}. If representational drift happens on a longer time scale than representation optimization, our formulation suggests that drift may be a consequence of changing environmental transition statistics (or a changing internal model of environmental statistics). This is somewhat compatible with other ideas about possible advantages of representational drift, e.g., for multi-task learning.


\clearpage

\bibliography{geobib}

\clearpage

\appendix

\section{Simplifying the objective}
\label{sec:app_simplify}

In this appendix, we will simplify the objective. Since the noise covariance matrix (see Sec. \ref{sec:math}) that defines internal state distances is diagonalizable, we have
\begin{equation}
\vec{\Sigma}^{-1} = \vec{Q}^T \vec{\Lambda}^{-1} \vec{Q} 
\end{equation}
and can make a change of variables to
\begin{equation}
\vec{r} = \vec{\Lambda}^{-1/2} \vec{Q} \vec{z} \ .
\end{equation}
In terms of this variable, the objective reads
\begin{equation}
\begin{split}
J[ \{ \vec{r}_x \}] &= - \sum_{x, y} p_0(x) p(y | x) \log p_{int}( y | x )  + \alpha \sum_x p_0(x) \frac{\Vert \vec{r}_x \Vert_2^2}{2}  
\end{split}
\end{equation}
where $p_{int}$ is now exclusively a function of the pairwise distances between the $\vec{r}_a$. 

We can reparameterize this objective in terms of a center $\langle \vec{r} \rangle$ and pairwise distances $d_{ij}$ between the $\vec{r}_a$. This is obvious for the Kullback-Leibler divergence term, so we only need to write the regularization term in terms of them. Note,
\begin{equation}
\Vert \vec{r}_x \Vert_2^2 = \Vert \vec{r}_x - \langle \vec{r} \rangle + \langle \vec{r} \rangle \Vert_2^2 = \Vert \vec{r}_x - \langle \vec{r} \rangle \Vert_2^2 + \Vert \langle \vec{r} \rangle \Vert_2^2 + 2 (\vec{r}_x - \langle \vec{r} \rangle) \cdot \langle \vec{r} \rangle \ .
\end{equation} 
Next,
\begin{equation}
\sum_x p_0(x) \frac{\Vert \vec{r}_x \Vert^2_2}{2}  = \sum_x p_0(x) \ \frac{1}{2} \left[ \ \Vert \vec{r}_x - \langle \vec{r} \rangle \Vert_2^2 + \Vert \langle \vec{r} \rangle \Vert_2^2 \ \right] \ .
\end{equation}

\noindent After some algebra, we can show
\begin{equation}
\Vert \vec{r}_x - \langle \vec{r} \rangle \Vert_2^2 = \sum_{a, b} p_0(a) p_0(b) \frac{1}{2} \left[ \ \Vert \vec{r}_x - \vec{r}_a \Vert_2^2 + \Vert \vec{r}_x - \vec{r}_b  \Vert_2^2 - \Vert \vec{r}_a - \vec{r}_b  \Vert_2^2  \  \right] \ .
\end{equation}
Finally,
\begin{equation}
\begin{split}
\sum_x p_0(x) \frac{\Vert \vec{r}_x \Vert^2_2}{2}  &= \sum_x p_0(x) \  \left[ \ \sum_{a,b} p_0(a) p_0(b) \left( \frac{d_{xa}^2}{4} + \frac{d_{xb}^2}{4} - \frac{d_{ab}^2}{4} \right) + \frac{\Vert \langle \vec{r} \rangle \Vert_2^2}{2} \ \right] \\
 &= \frac{\Vert \langle \vec{r} \rangle \Vert_2^2}{2} + \sum_{a, b} p_0(a) p_0(b) \frac{d_{ab}^2}{4} \ .
\end{split}
\end{equation}
Using this result, we get the reparameterized objective that appears in the main text.

\clearpage

\section{Details of uniform calculation}
\label{sec:app_uniform}

Consider a Markov chain for which $p_0(x) = 1/M$ for all $x$, and $p(y | x) = \frac{1 - \delta_{xy}}{M-1}$. This Markov chain is maximally symmetric, so our symmetric solution should have $d_{ij} = d$ for all $i, j \in \mathcal{X}$. This means
\begin{equation}
\begin{split}
q(\vec{r}_y | \vec{r}_x) &= \frac{1 - \delta_{xy}}{M-1} \\
p_e(\vec{r}_a | x) &= \frac{\delta_{ax} e^b + (1 - \delta_{ax}) e^{-d^2/2}}{e^b + (M-1) e^{-d^2/2}} \\
p_d(x | \vec{r}_a) &= p_e(\vec{r}_a | x) \ .
\end{split}
\end{equation}
Note that $Z(x) = Z$ is state-independent, and has value
\begin{equation}
Z = (M-1) e^{-d^2/2} \ .
\end{equation}
Multiplying out the encoding and decoding models, we have
\begin{equation*}
\begin{split}
p_{int}(y | x) &= \frac{1}{M-1} \sum_{a, b} p_e(\vec{r}_b | y) (1 - \delta_{ab}) p_e(\vec{r}_a | x) \\
&= \frac{1}{(M-1) (e^b + Z)^2} \sum_{a \neq b} \left[ \delta_{ax} e^b + (1 - \delta_{ax}) e^{-d^2/2} \right] \left[ \delta_{by} e^b + (1 - \delta_{by}) e^{-d^2/2} \right] \\
&= \frac{1}{(M-1) (e^b + Z)^2} \left\{ e^{2b} + 2 (M-2) e^{b - d^2/2} + \left[ M (M-1) - 2(M-2) + 1 \right] e^{-d^2} \right\} \\
&= \frac{1}{(M-1) (e^b + Z)^2} \left\{ e^{2b} + 2 (M-2) e^{b - d^2/2} + (M^2 - 3 M +5) e^{-d^2} \right\} \ .
\end{split}
\end{equation*}
Taking a logarithm,
\begin{equation*}
- \log p_{int} =  \log(M-1) + 2 \log(e^b + Z) - \log\left[ e^{2b} + 2 (M-2) e^{b - d^2/2} + (M^2 - 3 M +5) e^{-d^2}  \right] \ .
\end{equation*}
The objective becomes
\begin{equation*}
\begin{split}
J &= - \frac{1}{M(M-1)} \sum_{x \neq y} \log p_{int} + \alpha \frac{M(M-1)}{M^2} \frac{d^2}{4} \\
&= - \log p_{int} + \alpha \frac{M(M-1)}{M^2} \frac{d^2}{4} \\
&= \log(M-1) + 2 \log(e^b + Z) - \log\left[ e^{2b} + 2 (M-2) e^{b - d^2/2} + (M^2 - 3 M +5) e^{-d^2}  \right] + \alpha \frac{M(M-1)}{M^2} \frac{d^2}{4} \ .
\end{split}
\end{equation*}
Up to unimportant additive constants,
\begin{equation*}
J = 2 \log\left[ e^b + (M-1) e^{-d^2/2} \right] - \log\left[ e^{2b} + 2 (M-2) e^{b - d^2/2} + (M^2 - 3 M +5) e^{-d^2}  \right] + \alpha \frac{M-1}{M} \frac{d^2}{4} \ .
\end{equation*}

\clearpage

\section{Details of four state calculation}
\label{sec:app_four}

Consider a Markov chain for which $p_0(x) = 1/M$ for all $x$, and $p(y | x) = (\delta_{y,x+1} + \delta_{y,x-1})/2$ (where addition is done modulo $M$, although note that our state labels begin at $1$ rather than $0$). Here, we consider the case $M = 4$. The relevant symmetry group is $D_4$, the dihedral group of order $4$. Symmetry constraints tell us that $d_{12} = d_{23} = d_{34} = d_{41} =d$, and that $d_{13} = d_{24} = L$.

Relevant quantities include
\begin{equation}
\begin{split}
Z &= 2 e^{-d^2/2} + e^{-L^2/2} \\
q(\vec{r}_b | \vec{r}_a) &= \frac{(\delta_{b,a+1} + \delta_{b,a-1}) e^{-d^2/2} + \delta_{b,a+2} e^{-L^2/2}   }{Z} \\
p_e(\vec{r}_a | x) &= \frac{\delta_{ax} e^b + Z q(\vec{r}_a | \vec{r}_x)}{e^b + Z} \\
p_d(y | \vec{r}_b) &= p_e(\vec{r}_b | y) \ .
\end{split}
\end{equation}
Note also that all functions are symmetric, e.g., $q(\vec{r}_b | \vec{r}_a) = q(\vec{r}_a | \vec{r}_b)$. We can compute $p_{int}$:
\begin{equation*}
\begin{split}
p_{int} &= \sum_{a, b} p_e(\vec{r}_b | y) q(\vec{r}_b | \vec{r}_a) p_e(\vec{r}_a | x) \\
&= \frac{1}{(e^b + Z)^2} \sum_{a, b} \left[ \delta_{ax} e^b + Z q(\vec{r}_a | \vec{r}_x) \right] \left[ \delta_{by} e^b + Z q(\vec{r}_b | \vec{r}_y) \right] q(\vec{r}_b | \vec{r}_a) \\
&= \frac{1}{(e^b + Z)^2} \left\{ e^{2 b} q(\vec{r}_y | \vec{r}_x) + 2 Z e^b \sum_{a \neq x, y}  q(\vec{r}_y | \vec{r}_a) q(\vec{r}_a | \vec{r}_x) + Z^2 \sum_{a, b}  q(\vec{r}_y | \vec{r}_b) q(\vec{r}_b | \vec{r}_a) q(\vec{r}_a | \vec{r}_x)  \right\}  \ .
\end{split}
\end{equation*}
The above expression involves two- and three-step probabilities. These can be computed exactly in this case, although it is tedious. They technically only need to be computed (by symmetry) for one of the adjacent transitions (e.g., $1 \to 2$), since those are the only transitions that contribute to the objective function.

Let $1$ label the starting state, $2$ and $3$ be adjacent vertices, and $4$ be the farther vertex. The two-step probabilities are (9 relevant paths, quantified using $Z$):
\begin{equation}
\begin{split}
q^{(2)}(1 \to 1) &= q_{12} q_{21} + q_{13} q_{31} + q_{14} q_{41} = \frac{2 e^{-d^2} + e^{-L^2}}{Z^2} \\
q^{(2)}(1 \to 2) &= q_{13} q_{32} + q_{14} q_{42} = \frac{2 e^{-d^2/2 - L^2/2}}{Z^2} \\
q^{(2)}(1 \to 4) &= q_{12} q_{24} + q_{13} q_{34} = \frac{2 e^{-d^2}}{Z^2} \ .
\end{split}
\end{equation}
The relevant three-step probability is a sum of the probabilities of several paths:
\begin{equation}
\begin{split}
\frac{2 e^{ - \frac{3}{2} d^2} + e^{ - d^2/2 - L^2}}{Z^3} &= q_{12} \left[ q_{21} q_{12} + q_{23} q_{32} + q_{24} q_{42} \right] \\
\frac{2 e^{ - \frac{3}{2} d^2}}{Z^3} &= q_{13} \left[ q_{34} q_{42} + q_{31} q_{12} \right] \\
\frac{2 e^{ - d^2/2 - L^2} }{Z^3} &= q_{14} \left[ q_{43} q_{32} + q_{41} q_{12}  \right] \ . 
\end{split}
\end{equation}
Overall, the relevant three-step probability is
\begin{equation}
q^{(3)}(1 \to 2) = \frac{4 e^{ - \frac{3}{2} d^2} + 3 e^{ - d^2/2 - L^2}}{Z^3} \ .
\end{equation}
Finally, we can write that the relevant part of $p_{int}$ is
\begin{equation}
\begin{split}
p_{int} &= \frac{1}{(e^b + Z)^2 Z} \left\{ e^{2 b} e^{-d^2/2} + 2 e^b ( 2 e^{-d^2/2 - L^2/2}) + (4 e^{ - \frac{3}{2} d^2} + 3 e^{ - d^2/2 - L^2}) \right\}  \\
&= \frac{e^{-d^2/2}}{(e^b + Z)^2 Z} \left\{ e^{2 b}  + 4 e^b  e^{-L^2/2} + (4 e^{ - d^2} + 3 e^{  - L^2}) \right\}  \ .
\end{split}
\end{equation}
Taking a logarithm,
\begin{equation}
- \log p_{int} = \log Z + 2 \log(e^b + Z) + \frac{d^2}{2} - \log\left[  e^{2 b}  + 4 e^b  e^{-L^2/2} + 3 e^{  - L^2} + 4 e^{ - d^2}   \right] \ .
\end{equation}
The objective becomes
\begin{equation*}
\begin{split}
J &= \log Z + 2 \log(e^b + Z) + \frac{d^2}{2} - \log\left[  e^{2 b}  + 4 e^b  e^{-L^2/2} + 3 e^{  - L^2} + 4 e^{ - d^2}   \right] + \frac{\alpha}{4} \frac{2}{M^2} ( 4 d^2 + 2 L^2 ) \\
&= \log Z + 2 \log(e^b + Z) + \frac{d^2}{2} - \log\left[  e^{2 b}  + 4 e^b  e^{-L^2/2} + 3 e^{  - L^2} + 4 e^{ - d^2}   \right] + \frac{\alpha}{4} \left(  \frac{d^2}{2} + \frac{L^2}{4} \right) 
\end{split}
\end{equation*}
where the regularization term comes from counting the number of pairings of each kind.

\clearpage

\section{Details of general cyclic topology calculation}
\label{sec:app_cyclic}

As in the previous appendix, consider a Markov chain for which $p_0(x) = 1/M$ for all $x$, and $p(y | x) = (\delta_{y,x+1} + \delta_{y,x-1})/2$ (where addition is done modulo $M$). For simplicity, assume that $M$ is even, i.e., that $M = 2 N$ for some integer $N \geq 1$. Symmetry constraints tell us that distances should only depend on two vertices' relative positions along the `ring'. For example,
\begin{equation}
d_{12} = d_{23} = \cdots = d_{x,x+1}
\end{equation}
for any $x \in \mathcal{X}$. For $M$ even, there are $M/2$ unique distances that we must optimize (i.e., one hop away, two hops away, and so on); two vertices can only be at most $M/2$ edges apart. We will label these distances as $d_1$, $d_2$, ..., $d_{M/2}$.

Relevant quantities include
\begin{equation}
\begin{split}
Z &= e^{-d_{N}^2/2} + \sum_{k = 1}^{N-1} 2 e^{-d_k^2/2}  \\
q(\vec{r}_b | \vec{r}_a) &= \frac{\delta_{b,a+N} e^{-d_N^2/2} + \sum_{k=1}^{N-1} (\delta_{b,a+k} + \delta_{b,a-k}) e^{-d_k^2/2}    }{Z} \\
p_e(\vec{r}_a | x) &= \frac{\delta_{ax} e^b + Z q(\vec{r}_a | \vec{r}_x)}{e^b + Z} \\
p_d(y | \vec{r}_b) &= p_e(\vec{r}_b | y) \ .
\end{split}
\end{equation}
As in the previous appendix, we can write $p_{int}$ in terms of two- and three-step transition probabilities:
\begin{equation*}
\begin{split}
p_{int}(y | x) &= \frac{ e^{2 b} q(\vec{r}_y | \vec{r}_x) + 2 Z e^b \sum_{a \neq x, y}  q(\vec{r}_y | \vec{r}_a) q(\vec{r}_a | \vec{r}_x) + Z^2 \sum_{a, b}  q(\vec{r}_y | \vec{r}_b) q(\vec{r}_b | \vec{r}_a) q(\vec{r}_a | \vec{r}_x)    }{(e^b + Z)^2} \ .
\end{split}
\end{equation*}
The only difference is that these probabilities are now slightly more annoying to compute. Fortunately, only a single transition---the nearest neighbor transition, from any $x$ to $x+1$ (or equivalently, to $x-1$)---contributes to the objective, which by symmetry is equal to
\begin{equation}
\begin{split}
J &= - \log p_{int}(2 | 1) + \frac{2 \alpha}{4 M^2} \sum_{k=1}^{M-1} (M - k) d_k^2 \\
&= - \log p_{int}(2 | 1) + \frac{2 \alpha}{4 M} \left[ \frac{1}{2} d_N^2 +  \sum_{k=1}^{N-1} d_k^2 \right] 
\end{split}
\end{equation}
where $x = 1$ has been chosen arbitrarily, and where the details of the regularization term come from counting the pairwise distances of each kind. 

The second term in $p_{int} := p_{int}(2 | 1)$ involves $M - 2$ nonzero terms, and evaluates to
\begin{equation}
2 Z e^b \sum_{a \neq x, y} q(\vec{r}_y | \vec{r}_a) q(\vec{r}_a | \vec{r}_x) = \frac{4 e^b}{Z} \sum_{k=1}^{N-1} e^{-d_k^2/2 - d_{k+1}^2/2} \ .
\end{equation}
As mentioned in the main text, we will simplify this calculation by considering the $b \to \infty$ limit. In this limit, the three-step transition probability term can be ignored since it is of order $e^{- 2b}$. We have
\begin{equation*}
\begin{split}
p_{int} &= \frac{ q(\vec{r}_y | \vec{r}_x) + 2 Z e^{-b} \sum_{a \neq x, y}  q(\vec{r}_y | \vec{r}_a) q(\vec{r}_a | \vec{r}_x) + Z^2 e^{-2b} \sum_{a, b}  q(\vec{r}_y | \vec{r}_b) q(\vec{r}_b | \vec{r}_a) q(\vec{r}_a | \vec{r}_x)    }{(1 + Z e^{-b})^2} \\
&\approx \left[ q(\vec{r}_y | \vec{r}_x) + 2 Z e^{-b} \sum_{a \neq x, y}  q(\vec{r}_y | \vec{r}_a) q(\vec{r}_a | \vec{r}_x)  \right]    \left[ 1 - 2 Z e^{-b} \right] \\
&\approx q(\vec{r}_y | \vec{r}_x) + 2 Z e^{-b} \left\{ \sum_{a \neq x, y}  q(\vec{r}_y | \vec{r}_a) q(\vec{r}_a | \vec{r}_x)  -  q(\vec{r}_y | \vec{r}_x) \right]
\end{split}
\end{equation*}
to first order in $e^{-b}$. Explicitly, since only the $y = x+1$ term matters,
\begin{equation*}
\begin{split}
p_{int} &\approx \frac{e^{-d_1^2/2} + 2 e^{-b} \left[ 2 \sum_{k=1}^{N-1} e^{-d_k^2/2 - d_{k+1}^2/2}  -  Z e^{-d_1^2/2} \right]}{Z} \\
&= \frac{e^{-d_1^2/2}}{Z} \left\{ 1 + 2 e^{-b} \left[ 2 \sum_{k=1}^{N-1} e^{-d_k^2/2 - d_{k+1}^2/2 + d_1^2/2}  -  Z \right] \right\} \ .
\end{split}
\end{equation*}
We can now write
\begin{equation*}
\begin{split}
J &= \frac{d_1^2}{2} + \log Z - \log\left\{ 1 + 2 e^{-b} \left[ 2 \sum_{k=1}^{N-1} e^{-d_k^2/2 - d_{k+1}^2/2 + d_1^2/2}  -  Z \right]  \right\} + \frac{2 \alpha}{4 M} \left[ \frac{1}{2} d_N^2 +  \sum_{k=1}^{N-1} d_k^2 \right]  \\
&\approx \frac{d_1^2}{2} + \log\left[ e^{-d_{N}^2/2} + \sum_{k = 1}^{N-1} 2 e^{-d_k^2/2} \right] - 2 e^{-b} \left[ 2 \sum_{k=1}^{N-1} e^{-d_k^2/2 - d_{k+1}^2/2 + d_1^2/2}  -  Z \right]  + \frac{2 \alpha}{4 M} \left[ \frac{1}{2} d_N^2 +  \sum_{k=1}^{N-1} d_k^2 \right]  \ .
\end{split}
\end{equation*}

We would like to minimize this with respect to $d_1, ..., d_N$. The $d_1$ derivative is
\begin{equation}
\begin{split}
\frac{\partial J}{\partial d_1} &= d_1  - \frac{2 d_1 e^{-d_1^2/2}}{Z} - 4 e^{-b} d_1 \left[ \sum_{k=2}^{N-1} e^{\frac{-d_{k}^2 - d_{k+1}^2 + d_1^2}{2}} +e^{-d_1^2/2} \right] + \frac{\alpha}{M} d_1  \ .
\end{split}
\end{equation}
The derivative with respect to $d_k$ (for $2 \leq k < N$) is
\begin{equation}
\begin{split}
\frac{\partial J}{\partial d_k} &= - \frac{2 d_k e^{-d_k^2/2}}{Z} + 4 e^{-b} d_k \left[ e^{\frac{-d_{k-1}^2 - d_k^2 + d_1^2}{2}} + e^{\frac{-d_{k}^2 - d_{k+1}^2 + d_1^2}{2}} - e^{-d_k^2/2}\right] + \frac{\alpha}{M} d_k \ .
\end{split}
\end{equation}
The derivative with respect to $d_N$ is
\begin{equation}
\begin{split}
\frac{\partial J}{\partial d_N} &= - \frac{d_N e^{-d_N^2/2}}{Z} + 2 e^{-b} d_N \left[ 2 e^{\frac{-d_{N-1}^2 - d_N^2 + d_1^2}{2}} - e^{-d_N^2/2} \right] + \frac{\alpha}{2 M} d_N \ .
\end{split}
\end{equation}
Setting all of these equal to zero, but assuming all distances are nonzero, we have the following constraints:
\begin{equation}
\begin{split}
0 &= 1 - \frac{2 e^{-d_1^2/2}}{Z} - 4 e^{-b}  \left[ \sum_{k=2}^{N-1} e^{\frac{-d_{k}^2 - d_{k+1}^2 + d_1^2}{2}} +e^{-d_1^2/2} \right] + \frac{\alpha}{M}   \\
0 &= - \frac{2 e^{-d_k^2/2}}{Z} + 4 e^{-b}  \left[ e^{\frac{-d_{k-1}^2 - d_k^2 + d_1^2}{2}} + e^{\frac{-d_{k}^2 - d_{k+1}^2 + d_1^2}{2}} - e^{-d_k^2/2}\right] + \frac{\alpha}{M}  \\
0 &=- \frac{e^{-d_N^2/2}}{Z} + 2 e^{-b}  \left[ 2 e^{\frac{-d_{N-1}^2 - d_N^2 + d_1^2}{2}} - e^{-d_N^2/2} \right] + \frac{\alpha}{2 M}  \ .
\end{split}
\end{equation}
These equations in principle determine the $d_k$, but it does not appear possible to solve them analytically.

\end{document}